\documentstyle[aps,tighten]{revtex}

\begin{document}

\title{Cantor Spectra for Double Exchange Model}

\author{Atsuo Satou and Masanori Yamanaka}

\address{Department of Applied Physics, Science University of Tokyo, 
Kagurasaka, Shinjuku-ku, Tokyo 162-8601, Japan}

%\date{\today}

\maketitle

\begin{abstract}
We numerically study energy spectra and localization properties
of the double exchange model at irrational filling factor.
To obtain variational ground state, we use a mumerical technique
in momentum space by ``embedded'' boundary condition which has
no finite size effect a priori. 
Although the Hamiltonian has translation invariance,
the ground state spontaneously exhibits a self-similarity.
Scaling and multi-fractal analysis for the wave functions 
are performed and the scaling indices $\alpha$'s are obtained.
The energy spectrum is found to be a singular continuous, 
so-called the Cantor set with zero Lebesque measure.
\end{abstract}

\pacs{}

The study of 
the double-exchange (DE) model~\cite{REFde}
has a long history.
The purposes are to understand 
the electronic~\cite{REFdegennes} 
and transport properties~\cite{REFkubo}
of a class of the transition metal oxides,
mainly the perovskite manganites,
which are known to show the colossal magnetoresistance 
(CMR)~\cite{REFhelmolt}.
Recent experimental and theoretical studies have revealed
that it is a strongly-correlated quantum liquid 
in which many degrees of freedom are participating, such as
orbitals, lattice distorsion, electron correlation, dimensionality, 
the quantum nature of the spin, etc~\cite{REFtokura,REFtokuraJJAP}.
The current researches are in progress towards 
the investigation of the systems with many degrees of freedom.
A variety of ``double-exchange model''
have been introduced and investigated 
from many different viewpoints and approaches.

On the other hand,
the minimum model of the DE mechanism
has natures still to be clarified 
even before introduction of the many degrees of freedom.
(The model is composed of the itinerant $e_g$ electron 
and the localized $t_{2g}$ spin coupled 
with the Hund coupling each other.)
One is phase separation
~\cite{REFyunokidagotto,REFguinea,REFym,REFkkm,REFmyd,REFsarma}.
They has been found not only in the minimum model
also in those with direct exchange couplings 
between $t_{2g}$ spins~\cite{REFym} with orbital degree of freedom, 
and/or with electron correlation~\cite{REFmyd}.
Another is the spin-induced Peierls 
instability~\cite{REFykm,REFKoshibae}.
The ground state is characterized by a charge gap 
and a modulated spin structure.
The charge gap is induced by a commensurate modulation 
of the transfer integrals.
The spin state is distorted with a period
commensurate with the Fermi momentum to open the charge gap,
thus stabilizing the system.
This is isomorphic to the Peierls instability~\cite{REFpeielrs}
which has been discussed for a long time.
The only difference is that the spin degree of freedom is distorted,
instead of the lattice degree of freedom.

In this letter, we numerically study the electronic structure
of a one dimensional DE model 
at an irrational filling factor.
Although the Hamiltonian has a translation invariance,
the energy spectrum spontaneously exhibits 
a self similarity and forms the Cantor set.
The scaling analysis for the band width 
and the wave functions are performed.
We estimate the exponents $\alpha$'s
which characterize the localization properties~\cite{REFhk}.
To study the fine structures, usual boundary conditions,
such as periodic, twisted or open one, are useless. 
We use an alternative numerical technique in the momentum space 
with ``embedded'' boundary condition. 
This has no finite size effect a priori.
The localization properties and the similarity 
with the quasi-periodic system are discussed.
The electronic and transport properties
are found to be completely different from 
those of the DE model in the literatures. 

The DE model we employ is 
\begin{eqnarray}
H&=&-\Big(
\sum_{i=1}^{L}     t_{i,i+1}^{\phantom{\dagger}}
                   c_i^{\dagger}c_{i+1}^{\phantom{\dagger}}
+ \sum_{i=1}^{L/2} t_{2i-1,2i+1}^{\phantom{\dagger}}
                   c_{2i-1}^{\dagger}c_{2i+1}^{\phantom{\dagger}}
                   + h.c.\Big)
%\nonumber\\
%& &+J\sum_{\langle i,j \rangle}\vec{S_i}\cdot\vec{S_j}, 
+J\sum_{\langle i,j \rangle}\vec{S_i}\cdot\vec{S_j}, 
\label{eq:DE1}
\end{eqnarray}
where $c_{i,\sigma}$ is a fermion ($e_g$ electron) 
annihilation operator at site $i$,
$\vec{S}_i$ are localized ($t_{2g}$) spins 
which are treated as classical vectors directed along
$(\theta_i, \phi_i)$ in the spherical coordinates.
Moreover, $J$$>$0 is the direct exchange coupling strength 
between $t_{2g}$ spins.
We fix the parameter to be $JS^2/t$$=$1 throughout this paper. 
$L$ (even) is the total number of the sites.
The transfer integral~\cite{REFtransfer} is given by 
\begin{eqnarray} 
t_{i,j}=t\Big(\cos{\frac{\theta_i}{2}}
        \cos{\frac{\theta_j}{2}}
+e^{-i( \phi_i-\phi_j)} \sin{\frac{\theta_i}{2}}
                        \sin{\frac{\theta_j}{2}}\Big).
\label{eq:complexhoping}
\end{eqnarray} 
The pure one-dimensional version of the model (\ref{eq:DE1}),
i.e. in absence of the second term, 
all the sub-bands are dispersionless~\cite{REFKoshibae}.
We introduced the second term only as a perturbation to remove 
the singular behavior.
The singular behavior can be removed by any perturbation
and the form of the second term is not special 
to produce the band width.

We study incommensurate structure as a systematic limit of commensurate ones.
The difficulties we meet are the followings:
(i) The spin structure of the ground state is expected to be 
incommensurate, which are very sensitive to boundary conditions.
When we impose periodic, open, or twisted one,
the incommensurate structure is restricted to the corresponding space.
The true grournd state for a finite size does not necessarily exist
in the space.
(ii) The periodic boundary condition forces to align the 
spins periodically and spoils the fine structure.
One might expect that the incommensurate structure can be asymptotically 
embedded into the commensurate ones in a sufficiently large system.
However, the finite size correction is roughly estimated for $1/L$,
which is not small for our purpose.
The free boundary condition often induces a boundary state
and provides the similar problem.
(iii) We need a large denominator of the filling for the scaling analysis.
(iv) The ground state has a continuous infinite degeneracy.
To avoid these difficulties, we use ``embedded'' boundary condition 
to the spin part, 
The boudnary condition is similar to Ref.~\cite{REFKoshibae}.
We set the oriented relative angle 
between $\vec{S}_{L}$ and $\vec{S}_{L+1}$ is the same as
that between $\vec{S}_{0}$$=$$(0,0)$ and $\vec{S}_{1}$.
(The spins of $i$$>$$L$$+$$1$ are successively redefined 
from those of $1$$<$$i$$<$$L$$+$$1$
with maintaining the oriented relative angles.)
We impose the periodic boundary condition 
for the $e_g$ electron, i.e. $c_{L+1}$$=$$c_1$.
This enable us to treat the non-translation invariant spin states
under the Fourier transformation for the $e_g$ electron.
No finite size correction due to (i) and (ii) exists a priori.
We further assume the spin-induced Peierls 
conjecture~\cite{REFykm,REFKoshibae}.
Finally, at filling $x$$=$$p/(2q)$ the kinetic term of the Hamiltonian 
in the momentum space reads
\begin{eqnarray}
H_k(q;\{\phi_i,\theta_i\}) &=& \left(
\begin{array}{ccccccccc}
0&t_1e^{-ik}&t'_1e^{-2ik}&0&0&0& \cdots &t^{'*}_{2q-1}e^{2ik}&t^{*}_{2q}e^{ik}\\
t^{*}_1e^{ik}&0&t_2e^{-ik}&0&0&0& \cdots &0&0\\
t^{'*}_1e^{2ik}&t^{*}_2e^{ik}&0&t_3e^{-ik}&t'_3e^{-2ik}&0& \cdots &0&0\\
0&0&t^{*}_3e^{ik}&0&t_4e^{-ik}&0& \cdots &0&0\\
\multicolumn{9}{c}{\dotfill}\\
0& \cdots &0&t^{'*}_5e^{2ik}&t^{*}_{2q-4}e^{ik}&0&t_{2q-3}e^{-ik}&t'_{2q-3}e^{-2ik}&0\\
0& \cdots &0&0&0&t^{*}_{2q-3}e^{ik}&0&t_{2q-2}e^{-ik}&0\\
t'_{2q-1}e^{-2ik}& \cdots &0&0&0&t^{'*}_{2q-3}e^{2ik}&t^*_{2q-2}e^{ik}&0&t_{2q-1}e^{-ik}\\
t_{2q}e^{-ik}& \cdots &0&0&0&0&0&t^*_{2q-1}e^{ik}&0
\end{array}
\right)
\label{eq:matrix}
\end{eqnarray}
For brevity, we set $t_i$$\equiv t_{i,i+1}$ 
and $t'_{2i-1}$$\equiv t_{2i-1,2i+1}$.
To obtain the ground state, we optimize the functional
$E(\{\theta_i, \phi_i\})$$=$$E_t$$(\{\theta_i,$$\phi_i\})$
$+$$JS^2$$\sum_{\langle i,j \rangle}^{2q}$
$\big[\cos\theta_i\cos\theta_j$
$+$$\sin\theta_i$$\sin\theta_j$$\cos(\phi_i-\phi_j)\big]$
by the hybrid use of the conjugate gradient method 
and the simulated annealing by a monte-carlo method.
Here $E_t(\{\theta_i, \phi_i\})$ is
obtained by the numerical diagonalization of (\ref{eq:matrix}).
We should stress that the {\em infinite} system 
with {\em incommensurate} spin structure is 
embedded into the $2q\times2q$ matrix in the momentum space 
{\em without} finite size corrections.
The numbers of the parameters to be optimized are reduced to $4q$.

Our recursive procedure of arrival to the ground state 
are the followings:
(i) Set $q=2$.
(ii) Under the spin-induced Peierls conjecture,
the ground states of $x=p/q$ and $rp/(rq)$ are the same, 
where $p$, $q$, and $r$ are integers.
We optimize the ground state energy for the model 
with $q$-sites in a unit cell 
whose lowest $p$ sub-bands are filled.
(iii) We optimize the ground state energy for the model 
with $rq$-sites in a unit cell whose lowest $rp$ sub-bands are filled
so that the difference of the energy from (ii) 
below a desirable precision. (We set here three or four digits.)
(iv) Using the same procedure of (iii), 
we optimize the energy for $p/(rq)$-fillings
for each $q$, where $p$ and $rq$ are coprimers.
Then, we expect that the accuracy is the same 
that of (iii).
(v) Setting $q'$ to be a larger integer than $q$, return to (ii).
The procedure (i-v) also provides
a cross check of the spin-induced Peierls conjecture.   
Because this suggests the absence of a more larger
modulated structure than $q$-periodic one for $p/q$-filling.

We show the numerical results of the energy spectrum
in Fig.~\ref{fig:energyspec} 
where the fillings, $x=p/(2q)$, are exhausted all the combinations of 
$p=1$, $2$, $\cdots$, $17$ and 
$q=1$, $2$, $\cdots$, $18$.
A Peierls gap, which arised from the spin-induced Peierls mechanism,
opens at the Fermi surface. 
The structure of the energy spectrum is similar 
to that obtained by Hofstadter~\cite{REFhof}.
It is observed that the band structures
are different between two regions, $x<0.25$ and $0.25<x<0.5$.
In the former region, 
most of the sub-bands have enough dispersion 
and the energy gaps between the sub-bands are very small.
Most of the energy gaps cannot be displayed in this magnification.
In the latter, the sub-bands are less dispersive 
accompanied with larger energy gaps than those of the former.

To see these properties in detail,   
the electronic properties at the irrational filling 
are studied as systematic limits of rational ones.
To be specific, we focus on the cases $x^{(1)}$$=$$\varphi/4$ 
as a candidate for $x$$<$$0.25$
and $x^{(2)}$$=$$\varphi/2$ as that for $0.25$$<$$x$$<$$0.5$,
where $\varphi$$=$$(\sqrt{5}$$-$$1)/2$.
To do this, we use the Fibonacci numbers $F$, 
defined by $F_{n+1}$$=$$F_n$$+$$F_{n-1}$ and $F_0$$=$$F_1$$=$$1$.
Then the series of rational number $x_n^{(j)}$$=$$F_{n-1}/(k F_n)$
converges to $x^{(1)}$ for $k$$=$$4$ and $x^{(2)}$ for $k$$=$$2$
as $n$ becomes large.   
We denote the number of the sub-bands by $B_n$$\equiv$$kF_n$ (=$q$). 
In order to specify the branching of the energy bands,
we use sequence, $\{c_1,c_2,\cdots\}$~\cite{REFhk}.

The energy spectra of the series for $x^{(1)}$ is shown 
in Fig.~\ref{fig:energyseries}(a).
The energy gaps between the sub-bands are very small.
Except for the edge of the spectra,
the width of the sub-bands become uniform as the $n$ increases.
This suggests that rational $x$$=$$p/q$ gives bands with a width 
varying like $q^{-1}$ for large $q$.
The scaling analysis for the width, $W_n$, is shown in
the upper part of Fig.~\ref{fig:scaling}(a).
The states displayed in this figure are
(1) the first sub-band just below the fermi surface,
(or edge of the band defined by $\{0,1,1,1,\cdots\}$) and
(2) the center of the cluster shown in Fig.~\ref{fig:energyseries}(a)
(or center of the band, $\{0,0,0,0,\cdots\}$).
It is observed that $W_n$$B_n$$\sim$$1$ 
(i.e. $\alpha$$=$$1$) for (2),
while $W_n$$B_n$$\sim$$1/B_n$ (i.e. $\alpha$$=$$1/2$) for (1).
This means that all these states are extended.
At the band edge, $\alpha$ is $1/2$ when the state is extended.
This behavior comes from a remnant of van Hove singularities
in 1d bands.

In contrast to those of $x^{(1)}$,
the spectrum for $x^{(2)}$ forms clusters, as $n$ increases,
accompanied with various size of the energy gaps.
(See Fig.~\ref{fig:energyseries}(b).)
The middle part of the spectra for $n=5,6,7$ and $8$ are magnified 
and shown in the inset.
They show a striking similarity with those for $n=3,4,5$ and $6$.
This similarity is observed as well in the other regions.
(We should note that no self-similarity is found for $x^{(1)}$.
See Fig.~\ref{fig:energyseries}(a).)
The band width at the edges of the clusters seems to go down
faster than $q^{-1}$.
The scaling analysis for the width is shown the lower part
of Fig.~\ref{fig:scaling}(a).
The states displays in this figure are
the 1st ($\{0,1,1,1,\cdots\}$), 2nd, 3rd, and 4th sub-bands  
from the fermi surface below.
For the system with the structure shown 
in Fig.~\ref{fig:energyseries}(b),
we have to make a extrapolation in $n$ 
in every two or three points~\cite{REFhk}.
$W_n$ is observed to decrease rapidly.
The width goes down faster than an algebraic law
and is expected to be proportional to $exp(-\gamma B_n)$
where the $\gamma$ is a constant.
We have $\alpha=0$. Thus these states are localized.

The results of the wave function 
for series for $x^{(1)}$ and $x^{(2)}$ are shown 
in Figs.~\ref{fig:wavefunction}(a) and (b) columns, respectively. 
They are the wave functions for the energy band
just below the fermi surface.
For $x^{(1)}$, the localization length is likely proportional 
to the $q$, i.e. number of the sites in a unit cell.
The wave function for $x^{(2)}$ is localized.
These properties are consistent with the results from the analysis 
of the band width.
(However, the denominator of the filling is not large enough 
to make definitive conclusion only from Fig.~\ref{fig:wavefunction}(a).)

We perform the multi-fractal analysis for the wave function.
The method was established in~\cite{REFkadanoff,REFkohmoto}.
We use a more developed version~\cite{REFkoma}.
Extended, critical, and localized states are classified
by $\alpha_{min}=1$, $\ne 1,0$, and $=0$, respectively.
The estimates for the index $\alpha_{min}$ is shown 
in~\ref{fig:scaling}(b).
It is observed that the $\alpha_{min}$ for $x^{(1)}$
converges to a finite value.
This means that the state is not localized.
The $\alpha_{min}$ for $x^{(2)}$ is observed to
take a small value.
This means to the state has a tendency toward localization.
These results are consistent 
with the analysis for the energy spectrum 
and the behavior of the wave function shown 
in Fig.~\ref{fig:wavefunction}.

{\em Discussions:}
We have studied the energy spectra and wave function
of the DE model at zero temperature.
In studies of usual quasi-periodic systems, 
the quasi-periodicities are introduced by deterministic manners,
such as the on-site or off-diagonal quasi-periodicities.
In contrast to them, the self-similarity in the DE model
is spontaneously obtained from the translation-invariant 
Hamiltonian~\cite{REFssb}.
The modulation of the transfer integral is induced
from the spin-induced Peierls instability. 
In the every step of the iteration toward the irrational filling,
the modulation is expressed
by $q$ different amplitudes and, therefore, 
does not satisfy the Conway's theorem~\cite{REFconway}.
Therefore, the trace map would be described by a transcendant equation.

The numerical results suggest the existence of
two, at least, different insulating phases,
$x$$<$$0.25$ and $0.25$$<$$x$$<$$0.5$.
(The scaling analysis of band width, multi-fractal analysis for
the wave functions, and the features of the wave function and 
the energy spectra are mutually consistent.)
These phases can be distinguished by dynamical responses,
for example, the optical conductivity $\sigma(\omega)$.
Roughly estimated, the optical conductivity is characterized by the
the density of states near the Fermi level. 
The density of states for $x$$<$0.25 are very different from those 
for 0.25$<$$x$. For $x$$<$0.25 the energy gaps between the energy bands 
are infinitesimally small and the density of states exist except 
for the Peierls gap.
The optical conductivity is approximated by that of a usual 
band insulator (except for the contribution from the band edges of the
subbands).
On the other hand, for 0.25$<$$x$, the band structure exhibits a self-
similarlity. In the irrational limit, it is known that the 
optical conductivity is not well-defined in a usual sense~\cite{REFtakahashi} 
and is expected to maintain an anomalous scaling law.
(To study the exponents~\cite{REFtakahashi} is very interesting. 
However, even if we use the method proposed in this article the system size
and accuracy of the energy are too poor for such an analysis.)
The monte-carlo method can be applicable for systems in $d$$>$$1$.
In $d$$=$$2$, at {\it any} filling $x$, and for sufficiently large $J$,
it was conjectured~\cite{REFykm} that the ground state is a flux state 
whose energy spectrum is similar to the present one.
For several fillings $x$,
the conjecture is numerically confirmed~\cite{REFsy2},
because the phase factor in the transfer integral 
can open the Peierls gap~\cite{REFhlrw}.
However, a frustration between the incommensurate structure 
of the ground state and the periodicity of the periodic lattice 
arises in $d$$\ge$$2$.
Due to it, to obtain the fine structures of the system
are extremely difficult.

The approach in this paper would be 
one of the fairly good starting points 
for introducing other degrees of freedom,
such as the orbital~\cite{REFsy2}, lattice, electron correlation, 
quantum nature of the spins, anisotropies, etc.

The authors are grateful to Chisato Kanai, Tota Nakamura, 
Tatsuo Suzuki, and Kazuhiro Tada for the numerical correspondences,
and to Yoshiko Oi Nakamura for comments on the manuscript.
The authors thank the Supercomputer Center, 
Institute for Solid State Physics,
University of Tokyo for the use of the facilities.

\begin{figure}
\caption{
Energy spectrum near the Fermi surface as a function of filling $x$
at $JS^2/t=1$.
Spin-induced Peierls gap opens at the fermi surface.
We showed half bottom of the full spectrum.
For $x<0.25$, the energy bands are divided by small gaps.
The gaps are far smaller than the unevenness of carbon print 
and they look vanishing for the eyes.
}
\label{fig:energyspec}
\end{figure}

\begin{figure}
\caption{
Energy spectra below the fermi surface.
(a) Band structures 
for $x_n=1/8$, $2/12$, $3/20$, $5/32$, $8/52$, $13/84$, 
and $21/136$ are shown. This series converges to $x^{(1)}$.
Crosses indicate the location of the band edges.
The energy gaps between sub-bands are very small.
Therefore, the respective crosses which indicate the lower edge 
of an certain energy band and the upper edge of its lower band 
look to overlap.
(b) Band structures 
for $x_n=2/6$, $3/10$, $5/16$, $8/26$, $13/42$, $21/68$,
and $34/110$ are shown. The series converges to $x^{(2)}$.
The middle clusters for 
$8/26$, $13/42$, $21/68$, and $34/110$ are magnified 
and shown in the inset.
}
\label{fig:energyseries}
\end{figure}

\begin{figure}
\caption{
(a) Plots of $W_n B_n$ against $B_n$ for several states.
The upper three are for series $x^{(1)}$.
(Open squares and circles are, respectively, 
the 1st and 2nd sub-band from the fermi surface below. 
Crosses are for the band center.)
The lower four are for series $x^{(2)}$.
(Bold square, circle, triangle, and rombus are, respectively, 
the 1st, 2nd, 3rd, and 4th sub-bands 
from the fermi surface below.)
The index of the sequence of the energy bands are
$\{0,1,1,1,1,\cdots\}$ for the squares 
and $\{0,0,0,0,0,\cdots\}$ for the crosses.
(b) Estimates of the scaling index $\alpha_{min}$
for $x^{(1)}$ (open square) and $x^{(2)}$ (bold square)
as a function of $1/n$, where $n$ is the index of $B_n$.
}
\label{fig:scaling}
\end{figure}

\begin{figure}
\caption{
The wave functions $\vert \psi_i \vert^2$ of the energy band 
just below the fermi energy, (i.e. $\{0,1,1,1,1,\cdots\}$ series),
as a function of the site index $i$.
The left and right columns show those
for $x^{(1)}$ and $x^{(2)}$.
}
\label{fig:wavefunction}
\end{figure}


\begin{references}
\bibitem{REFssb}
We must be careful that in one dimension spontaneous breaking
of the continuous symmetry and that towards the incommensurate 
periodicity are forbidden.
\end{references}
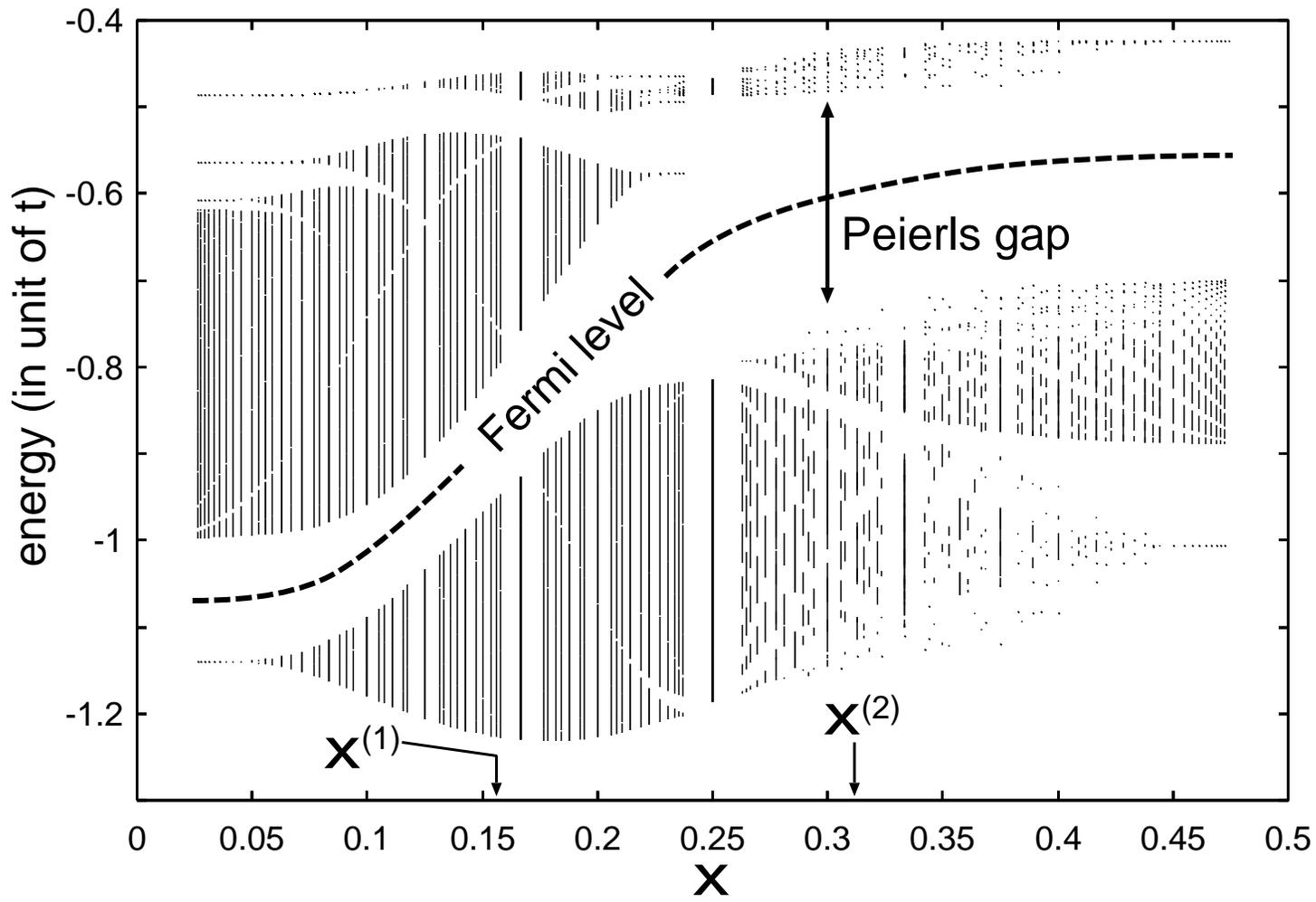
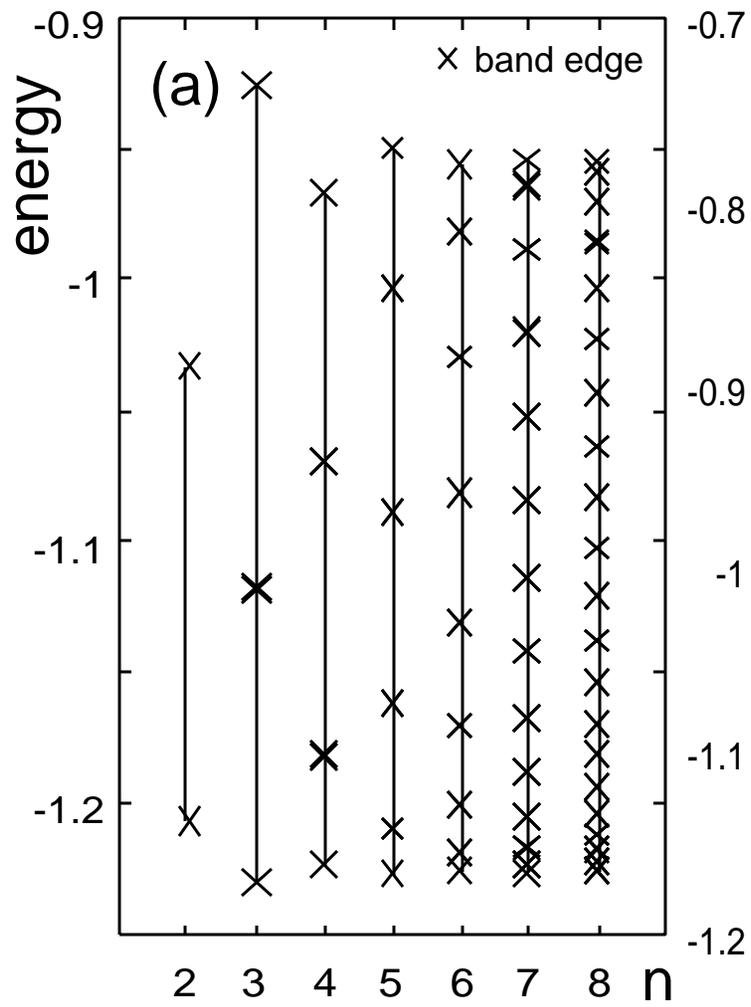
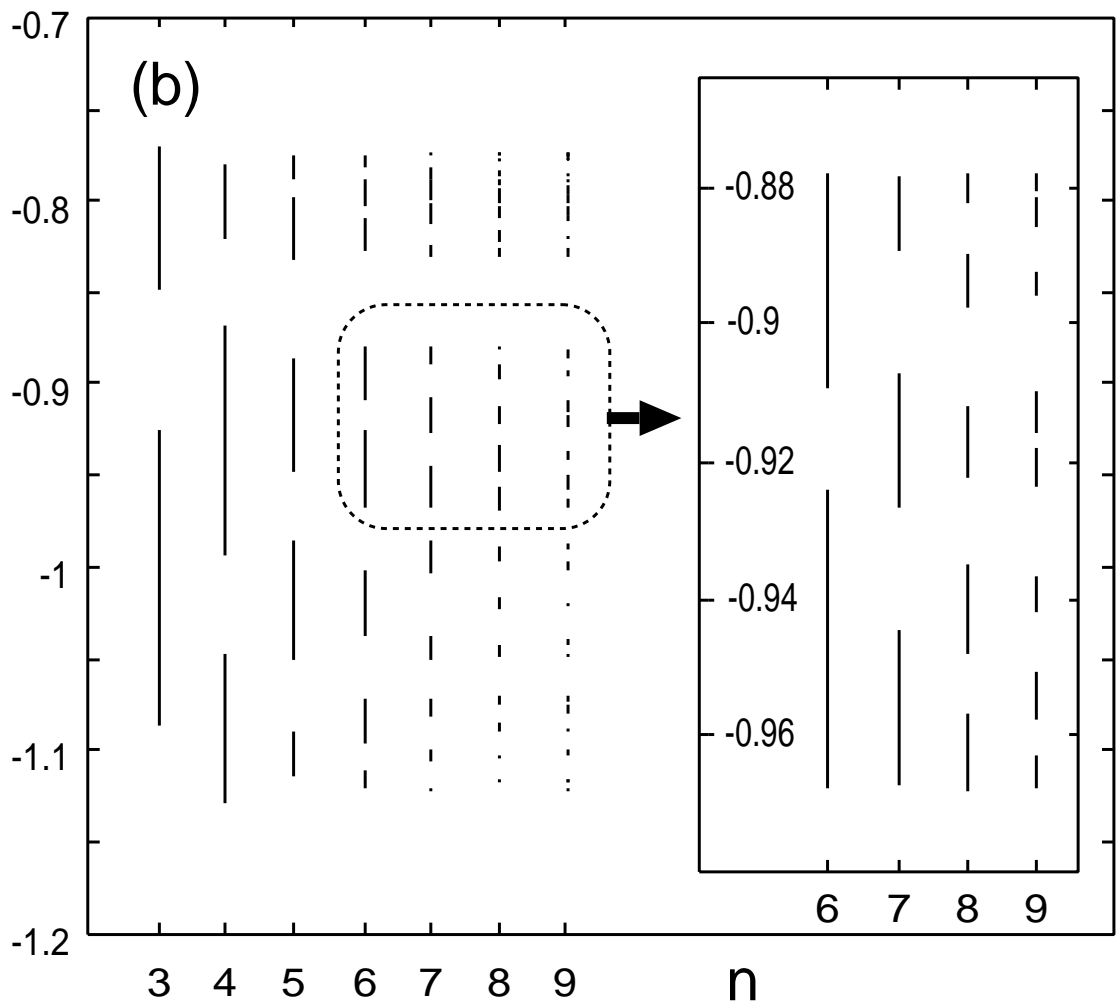
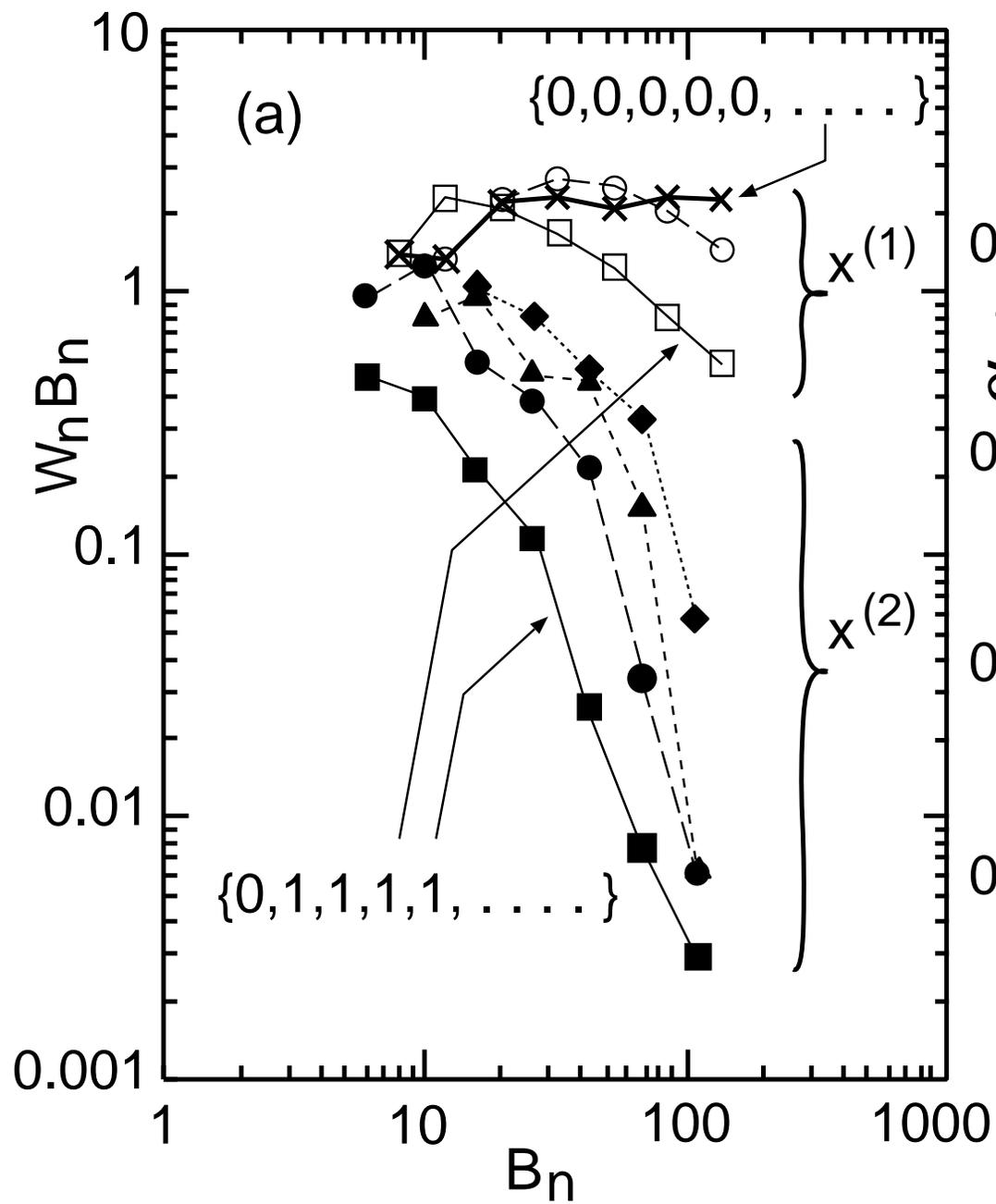
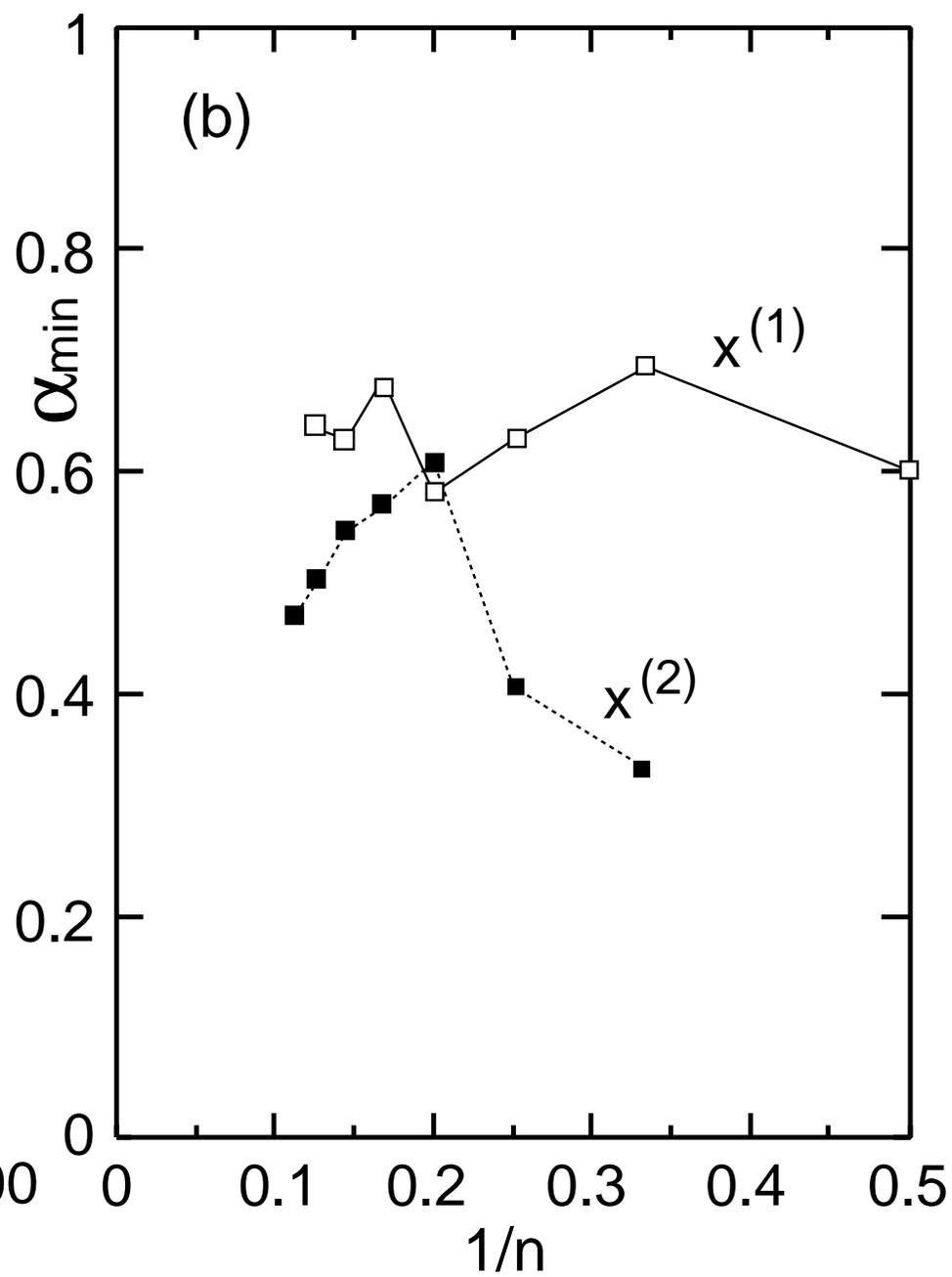
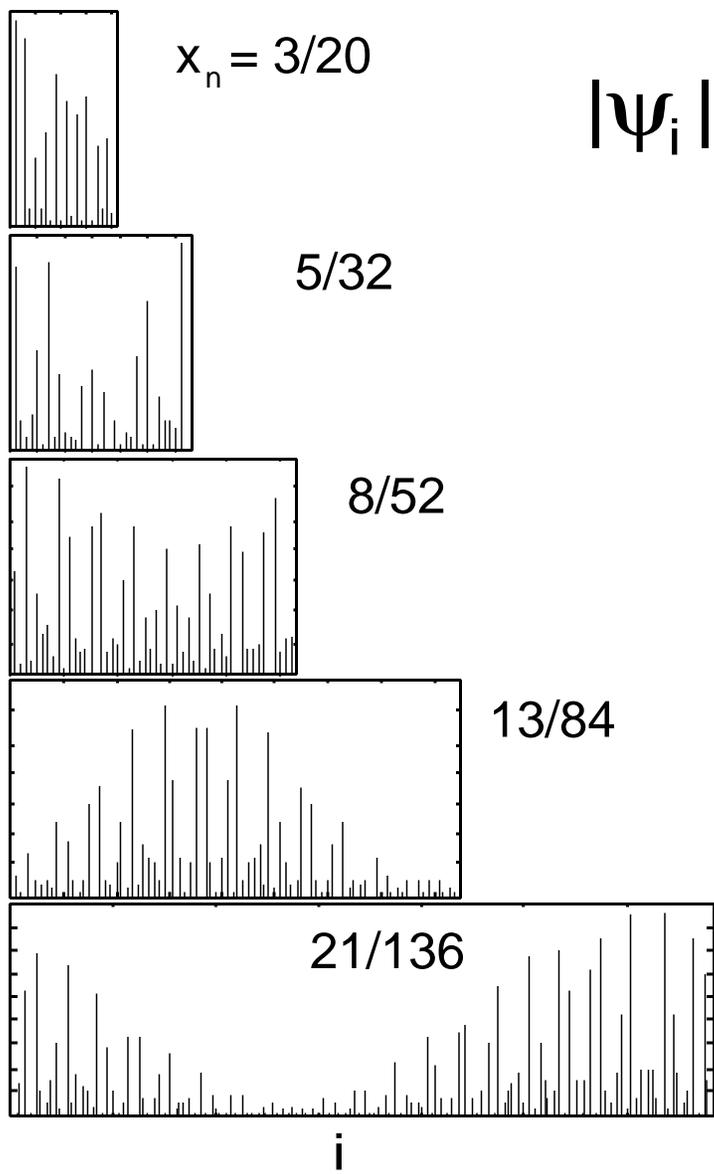
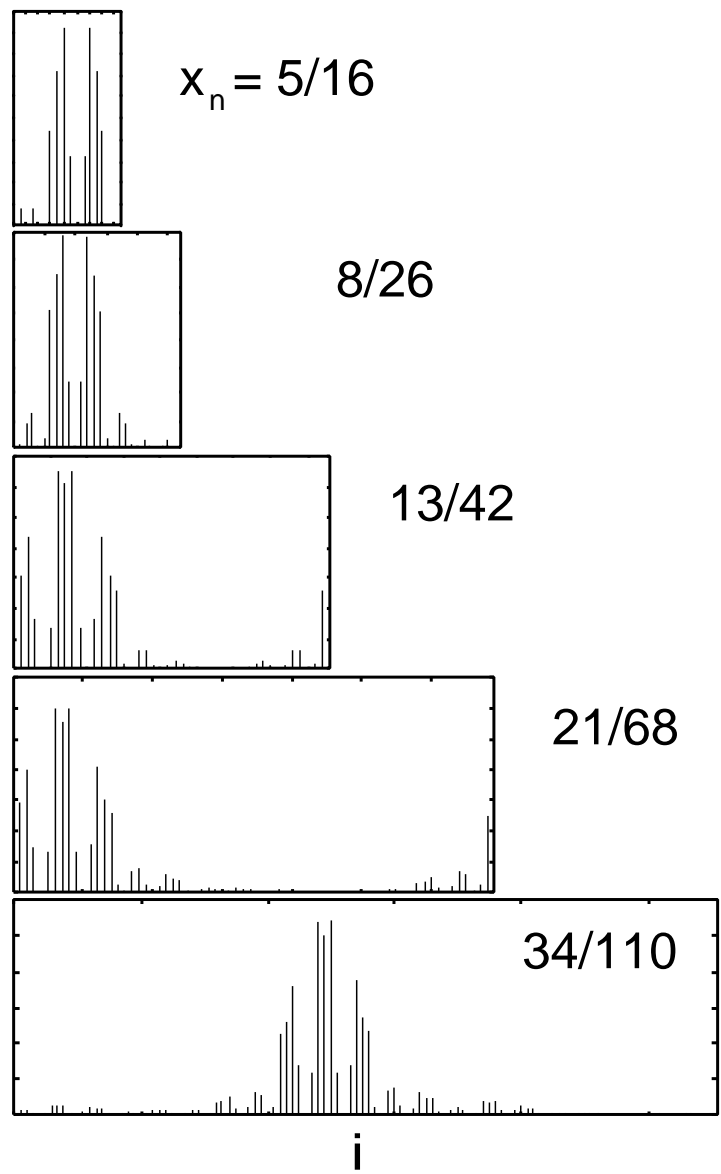
\end{document}